%% file: main.tex
\documentclass[journal=tches]{iacrtrans}

\usepackage{cite}
\usepackage{graphicx}
\usepackage{amsmath}
\usepackage{array}

\usepackage{wasysym}
\usepackage{enumitem}
\usepackage[]{todonotes}
\usepackage{algorithm} %
\usepackage{listings}
\usepackage[noend]{algpseudocode}
\algrenewcommand\algorithmicindent{0.75em} %
\usepackage{threeparttable}
\usepackage{booktabs}
\usepackage{hyperref}
\usepackage{multirow}
\usepackage{makecell}
\usepackage{amssymb}
\usepackage{bm}
\usepackage[table]{xcolor}
\usepackage{tikz}
\usetikzlibrary{arrows,arrows.meta,calc,fit,patterns,trees}
\usetikzlibrary{automata,positioning,matrix,backgrounds}
\usetikzlibrary{decorations.pathreplacing,angles,quotes}
\usetikzlibrary{shapes.symbols}

\usepackage{adjustbox}

\usepackage{colortbl}
\usepackage{xcolor}

\usepackage{standalone}
\usepackage[capitalise,noabbrev]{cleveref}

\usepackage{subcaption}
\usepackage{caption}
\usepackage[para]{footmisc}          %
\usepackage{minted}                %
\algrenewcommand\alglinenumber[1]{}

\newcommand{\schemename}{\textsc{TrEEStealer}}

\usepackage{xspace}

\newcommand{\ie}{i.e.,\xspace}
\newcommand{\eg}{e.g.,\xspace}

\newcommand{\etal}[1]{#1 \emph{et al.}}

\newcommand{\MyComment}[1]{\Comment{\textit{#1}}}

\usepackage{calc}    %
\makeatletter
\newcommand{\LineComment}[1]{%
  \Statex
  \begingroup
    \setlength\parindent{0pt}%
    \setlength\leftskip{\ALG@thistlm}%
    \noindent
    $\triangleright$\ \textit{#1}\par
  \endgroup
}
\makeatother

\definecolor{light_blue}{HTML}{77AADD}
\definecolor{orange}{HTML}{EE8866}
\definecolor{light_yellow}{HTML}{EEDD88}
\definecolor{pink}{HTML}{FFAABB}
\definecolor{light_cyan}{HTML}{99DDFF}
\definecolor{mint}{HTML}{44BB99}
\definecolor{pear}{HTML}{BBCC33}
\definecolor{olive}{HTML}{AAAA00}
\definecolor{pale_grey}{HTML}{DDDDDD}
\definecolor{light_grey}{HTML}{4f4f4f}

\begin{document}

\title{TrEEStealer: Stealing Decision Trees via Enclave Side Channels}

\author{Jonas Sander\inst{1} \and Anja Rabich\inst{1} \and Nick Mahling\inst{1} \and Felix Maurer\inst{1} \and Jonah Heller\inst{1} \and Qifan Wang\inst{2} \and Thomas Eisenbarth\inst{1} \and David Oswald\inst{2}}
\institute{
  University of Luebeck, Luebeck, Germany, \email{{j.sander,a.rabich,j.heller,thomas.eisenbarth}@uni-luebeck.de, nick.mahling@protonmail.com, maurerfelix@protonmail.com}
  \and
  Durham University, Durham, United Kingdom, \email{{qifan.wang2,david.f.oswald}@durham.ac.uk}
}

\maketitle
\keywords{Side Channel Attack \and Decision Tree \and Trusted Execution Environments}

\begin{abstract}
Today, machine learning is widely applied in sensitive, security-related, and financially lucrative applications. Model extraction attacks undermine current business models where a model owner sells model access, e.g., via MLaaS APIs. Additionally, stolen models can enable powerful white-box attacks, facilitating privacy attacks on sensitive training data, and model evasion.

In this paper, we focus on Decision Trees (DT), which are widely deployed in practice. Existing black-box extraction attacks for DTs are either query-intensive, make strong assumptions about the DT structure, or rely on rich API information. To limit attacks to the black-box setting, CPU vendors introduced Trusted Execution Environments (TEE) that use hardware-mechanisms to isolate workloads from external parties, e.g., MLaaS providers. We introduce \schemename{}, a high-fidelity extraction attack for stealing TEE-protected DTs. \schemename{} exploits TEE-specific side-channels to steal DTs efficiently and without strong assumptions about the API output or DT structure. The extraction efficacy stems from a novel algorithm that maximizes the information derived from each query by coupling Control-Flow Information (CFI) with passive information tracking. We use two primitives to acquire CFI: for AMD SEV, we follow previous work using the SEV-Step framework and performance counters. For Intel SGX, we reproduce prior findings on current Xeon 6 CPUs and construct a new primitive to efficiently extract the branch history of inference runs through the Branch-History-Register.

We found corresponding vulnerabilities in three popular libraries: OpenCV, mlpack, and emlearn. We show that \schemename{} achieves superior efficiency and extraction fidelity compared to prior attacks. Our work establishes a new state-of-the-art for DT extraction and confirms that TEEs fail to protect against control-flow leakage.
\end{abstract}

\section{Introduction}

\begin{figure}
    \centering    \includegraphics[width=0.45\linewidth]{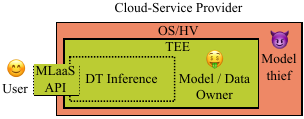}
    \caption{Attack scenario of \schemename{} in which a model owner provides a proprietary model protected through a TEE via a ML as a Service (MLaaS) API. The model thief respects the threat model of the TEE and just controls the operating system respective hypervisor, but still is able to steal whole real-world models precisely and efficiently.}
    \label{fig:setting}
\end{figure}

Model extraction attacks undermine the confidentiality of Machine Learning (ML) models, as they allow an attacker to extract models by repeatedly querying them in a black-box manner using only the model's outputs.
While much of the literature~\cite{wu2022model,jagielski2020high,liang2024model,DBLP:conf/uss/TramerZJRR16,chandrasekaran2020exploring} focuses on extracting neural networks, similar concerns apply to popular models such as Decision Trees (DTs), which are widely deployed in practice, e.g., in medical applications~\cite{curtis2025minimizing, roy2025decision, jiang2025urgency}, credit risk prediction~\cite{dumitrescu2022machine, chang2024credit} and intrusion detection~\cite{panigrahi2021consolidated}.
Being able to extract DTs allows recovery of feature value ranges along each path and tree structure, enabling downstream attacks that can infer properties of the training data~\cite{shokri2017membership,fredrikson2015model}.
Such leakage is undesirable as it may include sensitive information like medical records. Additionally, the models themselves are seen as intellectual property that should be kept secret.

In order to protect the confidentiality of such deployed models, developers increasingly host them in Trusted Execution Environments (TEEs) on cloud infrastructure~\cite{ohrimenko2016oblivious,law2020secure,wang2022enclavetree}. 
TEEs can protect individual processes (as Intel SGX) or entire Virtual Machines (VMs) (as Intel TDX and AMD SEV).
However, numerous works have demonstrated that TEEs remain vulnerable to a broad range of side channels~\cite{DBLP:conf/ches/MoghimiIE17, lee2017inferring, huo2020bluethunder, evtyushkin2018branchscope, DBLP:conf/ccs/BulckPS18, DBLP:conf/uss/MoghimiBHPS20, DBLP:journals/tches/WilkeWRE24}, which can ultimately break the confidentiality of the protected DT.
With these leakages, an attacker gains additional capabilities that can significantly strengthen DT extraction attacks.
Several prior works~\cite{DBLP:conf/uss/TramerZJRR16,chandrasekaran2020exploring,oksuz2024autolycus,wang2025barkbeetle} examined the extraction of DT models, though they operate in the general MLaaS setting and do not consider stronger scenarios where inference occurs inside a TEE.
Most of these~\cite{DBLP:conf/uss/TramerZJRR16,chandrasekaran2020exploring,oksuz2024autolycus}  rely primarily on black-box access to the model’s prediction interface and employ techniques such as identifying unique outputs, active query synthesis, or explainable Artificial Intelligence (xAI).
While applicable to TEEs in principle, these extraction methods suffer from limitations: dependence on rich output information, high query costs, and incomplete recovery of the tree structure.
Our evaluation shows that in real-world scenarios where such requirements are not satisfied, their extracted tree models often perform poorly or not at all. Additionally, they often recover only decision thresholds but miss repeated feature usage along a path, which reflects the feature’s relative importance and can leak information about the underlying training distribution.
Wang et al.\ additionally recover duplicate feature usage within each path~\cite{wang2025barkbeetle}, but rely on precise fault injection during inference.
Whether such techniques remain viable for TEE-protected DT inference is unclear, as injecting controlled faults at specific tree nodes is significantly more challenging within TEEs.

\paragraph{\textbf{Contributions}}
Motivated by the absence of comprehensive security evaluations for TEE-protected DT inference, we analyzed three popular DT inference libraries and found side-channel vulnerabilities in each of them. We demonstrate their exploitability through the introduction of \schemename{}, a high-fidelity DT extraction attack.
An overview of our approach is shown in \autoref{fig:setting}.
By using side-channel information from TEE execution to infer the DT branches while querying the MLaaS API, \schemename{} uses a novel algorithm to construct a shadow model of the DT. We demonstrate the practicality of our attack in case studies on two commercially available TEEs: Intel SGX and AMD SEV. We release our implementation of \schemename{} on GitHub\footnote{\href{https://github.com/UzL-ITS/TrEEStealer}{https://github.com/UzL-ITS/TrEEStealer}}.

\begin{itemize}[noitemsep,topsep=0pt]
    \item To the best of our knowledge, \schemename{} is the first attack that leverages microarchitectural primitives to extract DTs in TEE-protected inference.
    We introduce a new extraction algorithm with the lowest query complexity reported to date, enabling full reconstruction of the tree, including all features, thresholds, duplicate feature usage along each path, and internal node ordering.
    \item We identify vulnerable implementations in three widely used DT inference libraries, OpenCV, mlpack, and emlearn, which allow our attack to be mounted in realistic deployment settings.
    \item We demonstrate the effectiveness and practicality of \schemename{} for extracting DTs from Intel SGX and AMD SEV. For SGX, we exploit information from the Pattern History Register (PHR) to recover the branching behavior of the DT during its execution. For AMD Secure Encrypted Virtualization (SEV), we reconstruct the DT using a combination of a page-fault channel and reliable single-stepping.
\end{itemize}

\section{Preliminaries}
In the following, we introduce the preliminaries required to describe \schemename's end-to-end attack approach. \cref{tab:notation} reports the notations used in the explanation of \schemename's attack logic.

\subsection{Decision Trees}

\begin{table}[htbp]
    \centering
    \footnotesize
    \caption{Notations used in this paper. $(\bot)$ and $([\ldots])$ means the variable is initialized with bottom respective a list.}
    \label{tab:notation}
    \begin{adjustbox}{width=1.0\textwidth}
    \begin{tabular}{cl}
        \toprule
        \textbf{Notation} & \bf{Description} \\ 
        \midrule
		& \bf{Data}\\
		\cmidrule{2-2}
        \rowcolor{pale_grey}
        $m$ & Number of features\\
        $\mathbf{S}$ & Feature set $\mathbf{S}=[s_0,s_1,{\cdots},s_{d-1}]$ \\\rowcolor{pale_grey}
        $\mathbf{X}$ & Data sample $\mathbf{X}=[x_0,x_1,{\cdots},x_{d-1}]$ \\
        $\mathbf{R}^l, \mathbf{R}^u$ & \makecell[l]{Feature ranges $\mathbf{R}=[r_0, r_1, {\cdots}, r_{d-1}]$ with lower $r_i^l$ and upper limit $r_i^u$ per feature $s_i$}\\
        \midrule
		& \bf{(Shadow) Decision Tree Node}\\
		\cmidrule{2-2}
		$\epsilon$ & Extraction resolution of node thresholds\\\rowcolor{pale_grey}
        $\mathcal{V}$ & Decision tree node $\mathcal{V}=\operatorname{Node}(p,d,\mathbf{X},\mathbf{B})$\\
        $\mathcal{V}.p$ & Parent of the node\\\rowcolor{pale_grey}
        $\mathcal{V}.l$ & Left child of the node $(\bot)$\\
        $\mathcal{V}.r$ & Right child of the node $(\bot)$\\\rowcolor{pale_grey}
        $\mathcal{V}.d$ & Depth of the node \\
        $\mathcal{V}.f$ & Feature of the node $(\bot)$\\\rowcolor{pale_grey}
        $\mathcal{V}.t$ & Threshold of the node $(\bot)$\\
        $\mathcal{V}.v$ & Value (label) if the node is a leaf $(\bot)$\\\rowcolor{pale_grey}
        $\mathcal{V}.{\mathbf{X}}$ & Input that initially explored the node\\
        $\mathcal{V}.{\mathbf{B}}$ & \makecell[l]{Branching information$[b_i | b_i \gets 0\text{ if }\mathcal{V}.\mathbf{X}_i\text{ led to left traversal, else } b_i \gets 1]$} \\\rowcolor{pale_grey}
        \colorbox{pale_grey}{\makecell{$\mathcal{V}.{\mathbf{T}^0}$,\\$\mathcal{V}.{\mathbf{T}^1}$}} & \makecell[l]{Threshold ranges - Minimal respectively maximal values which resulted in a left (0)\\ respectively right (1) classification $([])$}\\
        \makecell{$\mathcal{V}.{\mathbf{DD}},$\\$\mathcal{V}.{\mathbf{TT}}$}  & \makecell[l]{Feature thresholds - Feature-indexed lists of lists of thresholds ($\mathbf{TT}$) and\\corresponding node depths ($\mathbf{DD}$) of already confirmed thresholds on the path to\\ node $\mathcal{V}$($[l|l\gets[]\text{ for }i\gets 0, i<m,i\gets i+1],[l|l\gets[]\text{ for }i\gets 0, i<m,i\gets i+1]$)}\\
        \midrule
        & \bf{(Shadow) Decision Tree Model} \\
        \cmidrule{2-2}
        $\mathcal{T}$ & Decision Tree $\mathcal{T}=\operatorname{DT}(r, \mathbf{I})$, $\mathcal{\tilde{T}}$ for the shadow DT\\\rowcolor{pale_grey}
        $\mathcal{T}.r$ & Root node\\
        $\mathcal{\tilde{T}}.{\mathbf{I}}$ & Backlog of incomplete nodes\\\rowcolor{pale_grey}
        $\mathcal{T}(\mathbf{X})$ & Perform inference using DT $\mathcal{T}$ and input $\mathbf{X}$\\
        \bottomrule
    \end{tabular}
    \end{adjustbox}
\end{table}

\begin{figure}
    \centering    \includegraphics[width=0.65\linewidth]{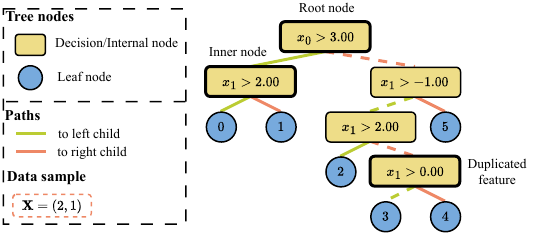}
    \caption{Target DT with rectangles representing decision nodes (i.e., internal nodes). The circles represent the leaves containing class labels (for classification) or numerical predictions (for regression). The dashed line visualizes the inference process. Representative nodes for explanation of the extraction process are highlighted.}
    \label{fig:target_dt}
\end{figure}

DTs are a class of supervised learning models widely used for classification and regression tasks due to their interpretability, low computational cost, and robustness to heterogeneous feature types.
Without loss of generality, we consider DTs with a binary structure, meaning each node has either a left and a right child or is a leaf node containing the final classification or regression value (see also \cref{fig:target_dt} and \cref{tab:notation}).
Additionally, each node has exactly one parent, except for the root node, which has no parent. During the inference process, an input $\mathbf{X}$ is processed starting in the root node. Each inner node checks one feature value against a threshold value $x_i>\mathcal{V}.t$. If the check evaluates to true, the inference proceeds along the left child and otherwise along the right child. The process terminates in a leaf node and returns its classification or regression value. The structure of the DT, including the threshold values, is determined in the corresponding training process.

\subsection{Trusted Execution Environments}

TEEs combine hardware-enforced isolation with Remote Attestation (RA), which allows remote parties to verify the integrity of enclave code and makes TEEs particularly attractive for confidential cloud computing such as MLaaS.
TEEs can be used for applications (\eg Intel Software Guard Extensions (SGX)) or for entire VMs (\eg SEV). In this work, we consider both SGX and SEV as our target scenarios, each utilizing different attack surfaces. The following section provides an overview of them.

\noindent\textbf{Intel SGX} allows the hardware-supported isolation of applications in their own memory regions called enclaves, which are encrypted in DRAM. This prevents access to the code and data by the Operating System (OS) and HV, while also allowing the enclave to use system resources, such as memory. The SGX Trusted Computing Base (TCB) assumes that only the CPU supplied by Intel is trusted. When code is executed by the CPU, the MMU will decrypt relevant code and data pages.

\noindent\textbf{AMD SEV} allows running confidential VM guests that are isolated from the hypervisor. Pages belonging to SEV VMs are stored encrypted in the DRAM and decrypted by the MMU. Each VM has its own encryption key, which is managed by the AMD Secure Processor (PSP). SEV Encrypted State (SEV-ES) is an extension that introduces the VM Save Area (VMSA), for encrypting the register states of the VM before storing them in the VM Control Block (VMCB). Secure Nested Paging (SNP) further adds a shadow table for storing the ownership of pages used by a guest, the Reverse Map Table (RMP). This ensures memory integrity by preventing memory replay and re-mapping attacks by the hypervisor. With SEV, we are referring to SNP unless explicitly stated otherwise.

\subsection{Performance Counters}

Hardware Performance Counters (HPC) allow for profiling applications based on events and cycles. The exact supported events are vendor-specific and may vary based on the exact CPU. HPCs are programmed and read using dedicated Model Specific Registers (MSRs). As they provide valuable information on executed code and data used, they can also reveal sensitive information of workloads. Hence, they are not accessible to the host OS for SGX unless the enclave is built as a debug enclave (our threat model does not assume debug enclaves). However, SEV allows the hypervisor to access and configure a guest TEE performance counter events up to Zen 5.

\subsection{Conditional Branch Prediction}
The Conditional Branch Predictor (CBP) minimizes pipeline stalls by predicting the outcome of conditional branches (taken or not-taken) before their actual resolution. These predictors have evolved from simple local designs to sophisticated TAgged GEometric history length (TAGE) architectures that effectively mitigate aliasing effects \cite{DBLP:journals/jilp/Seznec07, mahling2023reverse, yavarzadeh2023half}. Aliasing means two different branches modify the same entry in the so-called Branch History Table resulting in decreased prediction performance. Modern Intel processors implement a variant of the TAGE predictor, as demonstrated through reverse engineering efforts. This implementation comprises two fundamental components: the Pattern History Table (PHT), which maintains predictions for individual branches, and the PHR, which captures the execution history of recent branches \cite{yavarzadeh2023half,mahling2023reverse,yavarzadeh2024pathfinder}.

\noindent\textbf{Pattern History Register.} The PHR functions as a global shift register within each CPU core, maintaining information from recently executed taken branches \cite{mahling2023reverse, yavarzadeh2023half}. Branches that are not taken leave the PHR unchanged. The register operates using 2-bit units called doublets \cite{yavarzadeh2024pathfinder}. When a branch is taken, the PHR shifts by one doublet and incorporates a footprint derived from both the branch address and its target address through an XOR operation.
With a capacity of 194 doublets, the PHR preserves information from the last 194 taken branches.

\noindent\textbf{Pattern History Table.} Modern Intel processors implement the PHT as a table of 3-bit saturating counters that are indexed during branch prediction \cite{mahling2023reverse}. Within the TAGE-like predictor on Intel, multiple PHTs exist: a base table indexed exclusively by address bits, and three tagged tables that incorporate the PHR to consider the execution history of previous branches when predicting the current conditional branch. This design enhances prediction accuracy for branches whose outcomes correlate with historical execution patterns. For the tagged tables, the index calculation combines the 6th bit of the branch address with a 7-bit sequence derived from folding portions of the PHR. Each of the three tables uses progressively longer parts of the PHR. To reduce collisions of multiple branches, the PHR has tags similar to CPU caches. These tags comprise the 13 least significant bits of the branch address. When multiple PHTs yield valid predictions due to matching tags across several tables, the predictor selects the prediction from the PHT that incorporates the longest history \cite{yavarzadeh2023half}.

\subsection{TEE-based Side-Channels}

\noindent\textbf{Controlled-channel and Address Translation-based Attacks.} Controlled-channel attacks, a concept introduced by\cite{DBLP:conf/sp/XuCP15}, exploits the privileges of the OS to monitor and manipulate page tables and thereby observe enclave memory-access patterns (see also \cite{shinde2016preventing}). 
Although SGX enclaves protect code and data from a compromised OS, the OS still controls virtual memory and can use page faults or page-table metadata to infer program behaviour.

\noindent\textbf{Interrupt-driven Attacks.} Interrupt-driven frameworks provide fine-grained control over enclave or VM execution. 
SGX-Step \cite{DBLP:conf/sosp/BulckPS17} uses precise interrupt delivery (e.g., via APIC timers) to implement deterministic single-stepping of enclave code, which enables detailed monitoring of memory accesses over time and supports controlled-channel exploitation. CopyCat combines the use of a page-fault controlled channel to track when the enclave accesses certain code and data pages and single-stepping to reconstruct the exact control-flow path executed by the enclave~\cite{DBLP:conf/uss/MoghimiBHPS20}. In the VM setting, SEV-Step \cite{DBLP:journals/tches/WilkeWRE24} implements a similar single-stepping capability for SEV guests and exposes primitives for page-fault tracking and cache attacks against SEV-protected VMs.

\noindent\textbf{Branch-prediction Attacks.} Branch-prediction attacks target the CPU's prediction structures to learn control-flow decisions of a victim. 
They include attacks on the branch target buffer (BTB) \cite{lee2017inferring,chen2019sgxpectre}, the pattern history table (PHT) \cite{evtyushkin2018branchscope,huo2020bluethunder}, and the return stack buffer (RSB) \cite{koruyeh2018spectre}. 
BTB-based techniques detect whether a specific branch target has been trained by the victim.
PHT-style attacks induce collisions between attacker and victim branches to infer branch direction.
RSB attacks poison the return stack to trigger misspeculation and subsequent leakage.

\section{\schemename}

In the following, we introduce the threat model and give a high-level overview of the attack to finally introduce each component in detail.

\subsection{Threat Model}
We assume the attacker can query the DT model via standard MLaaS APIs, receiving only minimal outputs (e.g., classification/regression results without confidence values). Incomplete queries\footnote{Supported by some MLaaS platforms (\eg BigML), where leaving a feature unspecified causes traversal to stop at a split on that feature, directly revealing its threshold.} are disallowed, preventing direct leakage of internal node information. The attacker aims to reconstruct the full DT structure (feature assignments and thresholds) with minimal queries, without prior knowledge of the tree. However, they know feature value ranges (commonly documented or derivable from semantics, e.g., human height less than 5m). In addition, we assume the standard threat model for TEEs: for Intel SGX, the attacker controls the OS (the enclave is not built in debug mode); for AMD SEV, they hold hypervisor privileges. We emphasize that these are not specific assumptions of our work; control over interrupts (enabling single-stepping) and access to microarchitectural state (like the PHR) are established capabilities for such privileged attackers. Specifically, they allow: (1) programming APIC timers for precise one-shot interrupts, and (2) manipulating page tables to infer code execution locations at page granularity. The target model is pretrained, and training details are outside our scope.

\subsection{Attack Overview}
\begin{figure}
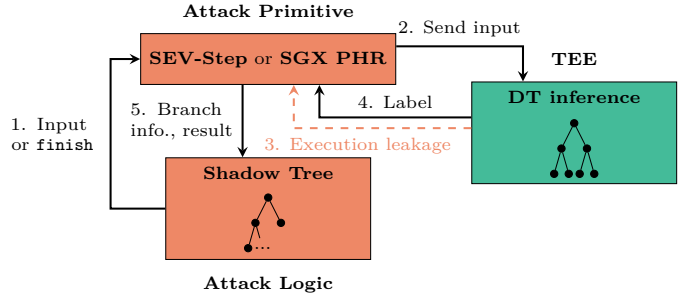

    \centering
    \includestandalone[width=0.65\columnwidth]{figs/treestealer_diagram}
    \caption{Overview of \schemename's attack workflow.}
    \label{fig:attack_workflow}
\end{figure}

From a high-level perspective, \schemename{} consists of two modules: a target dependent attack primitive and the attack logic. The attack primitive exploits side-channels of the targeted TEE to extract branching information from the inference process. The attack logic generates inputs to perform the inference and leverages the extracted branching information to reconstruct the DT model under attack (see also \cref{fig:attack_workflow}).
The attack performs the following basic steps:
\begin{enumerate}[noitemsep,leftmargin=*,topsep=0pt]
    \item The attack primitive gets an input from the attack logic.
    \item The attack primitive sends the input via a MLaaS API to the target TEE running a DT inference service.
    \item The TEE starts the inference process and the attack primitive monitors the branching decisions of the inference (left and right decisions in the DT structure).
    \item The TEE sends the inference result via the MLaaS API back to the attack primitive.
    \item Given the inference result and branching information the attack logic builds a shadow version of the target model by iteratively adjusting the inputs to map features to nodes, find threshold values, and traverse all paths to extract the whole DT. While the shadow DT is incomplete the attack logic returns to step 1 and sends the adjusted input to the attack primitive, otherwise the attack logic terminates the attack by sending \texttt{finish}.
\end{enumerate}

\subsection{Attack Logic}

Below, we note lists with bold capital letters, e.g., $\mathbf{X}$ and structs with calligraphic letters, e.g., $\mathcal{V}$. List elements and struct members are accessed via the $(.)$-operator and their index respective member name. $\text{List}.\operatorname{append}(x)$ appends an element $x$ to the end of the list. $\text{List}.\operatorname{last}()$ returns and $\text{List}.\operatorname{pop}()$ removes and returns the last element of a list. $\text{List}.\operatorname{empty}()$ returns $\top$ if the list is empty and $\bot$ else. $\operatorname{min}(\text{list})$ and $\operatorname{max}(\text{list})$ return the minimum respective maximum of a given list and $\operatorname{len}(\text{list})$ the number of elements in the list. Given two lists $\mathbf{L}$ and $\mathbf{LL}$, with $n=\operatorname{min}(\operatorname{len}(\mathbf{L}), \operatorname{len}(\mathbf{LL}))$ $\operatorname{zip}(\mathbf{L}, \mathbf{LL})$ returns an iterator of tuples $(\mathbf{L}.0, \mathbf{LL}.0), (\mathbf{L}.1, \mathbf{LL}.1), \ldots, (\mathbf{L}.n, \mathbf{LL}.n)$.	Structs and lists are passed as references and $\operatorname{Copy}()$ performs a deepcopy (copy by value).

In the following, we introduce \schemename's approach to explore all nodes of the private target DT while limiting the number of required queries through passive information tracking. We start with a high-level explanation to give an intuitive understanding of the algorithm. Then, we explain the core extraction logic based on three exemplary situations in the extraction process of the target tree $\mathcal{T}$ in \cref{fig:target_dt} in detail. We start with the initial situation of an empty shadow tree $\mathcal{\tilde{T}}$ and show how the complete extraction of the root node $\mathcal{\tilde{T}}.r$ is performed. Then, we demonstrate the extraction of a deeper node, and finally, we set out how the extraction of nodes with duplicated features along their inference path is realized. For our example, we set the extraction resolution for the thresholds $\epsilon=0.5$. Similar to previous work, $\epsilon$ will be experimentally tuned depending on the target datasets.

\newcommand*\mcirc[1]{%
  \tikz[baseline=(char.base)]\node[shape=circle,draw,inner sep=0.8pt] (char) {$\scriptstyle #1$};}

\subsubsection{High-level Explanation}
In this section, we present the high-level ideas of the extraction logic starting with the initial input and how the node exploration including information tracking is performed. We then give an intuitive explanation of the feature- and threshold-extraction process.

\noindent\textbf{Initial Input and Node Exploration.} Based on the given feature ranges $\mathbf{R}^l, \mathbf{R}^u$, the attack logic generates an initial input that traverses the left-most path of the target DT. For each input  the attack primitive tracks the corresponding branching information of the target DT and the attack logic leverages them to add empty nodes (without feature and threshold values) to an empty shadow tree. The leaf nodes of the shadow DT are implicitly given through the inference results. Additionally, each new empty node is added to a first-in-first-out (FiFo) backlog of incomplete target nodes for feature- and threshold-extraction in the next steps. Since parents are always added before their children, the FiFo principle of the backlog ensures that the extraction of a node does not begin before the extraction of all of its ancestors is completed, enforcing a top-to-bottom extraction.
As described in the following, the feature- and threshold-extraction enforces left-right-traversals at the currently targeted node, ensuring the successive exploration of the whole tree structure.

\noindent\textbf{Passive Information Tracking.} To reduce the overall number of queried inputs, the attack logic passively tracks threshold-bounds of all nodes along the inference path during each inference run.
Each node contains two lists $\mathcal{V}.{\mathbf{T}^0}, \mathcal{V}.{\mathbf{T}^1}$ we call threshold ranges which track the minimal input value resulting in a left- and the maximal input value resulting in a right-traversal.
Additionally each node contains two feature-indexed list of lists $\mathcal{V}.\mathbf{TT}, \mathcal{V}.\mathbf{DD}$ we call feature thresholds which track the threshold values and the corresponding node depths of all nodes on the path to $\mathcal{V}$ that check on a specific feature.

\noindent\textbf{Feature- and Threshold-Extraction.} After the initial exploration of the left-most path, the attack logic iteratively crafts inputs to extract the feature and threshold of the next empty node in the backlog and adds the collected information to the shadow DT. To determine the feature of a node, the attack logic leverages the input that initially explored the node and toggles one feature after another until the branching behavior of the node diverges. If another node on the path to the currently targeted node already checks the toggled feature we use the tracked feature thresholds to ensure that the new crafted input still reaches the currently targeted node.
To determine the threshold of a node, the attack logic performs a binary search based on the tracked threshold ranges.

\subsubsection{Initialization}
\begin{figure}[t!]
    \centering    \includegraphics[width=0.175\linewidth]{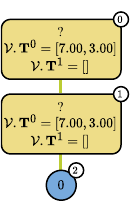}
    \caption{Shadow DT after the first crafted input and with per-node extracted threshold ranges. The circled numbers show the order in which the nodes were explored and how they are added to the backlog of incomplete nodes $\mathcal{\tilde{T}}.\mathbf{I}$.}
    \label{init_extraction}
\end{figure}

Given the ranges of each feature with lower limits $\mathbf{R}^l=[2, -2]$ and upper limits $\mathbf{R}^u=[7, 3]$, the attack logic starts by generating the initial input $\mathbf{X}=[7, 3]$, which traverses the leftmost path of the target DT (see \cref{fig:target_dt} and the outer attack loop in \cref{algo:dtextraction}). Corresponding to the inference run, the attack logic receives the inference result $v=0$ and the branching information $\mathbf{B}=[0,0]$ from the attack primitive $\operatorname{AttackPrim[]}$. Note, a left traversal is represented by the 0- and the right traversal by the 1-bit. Based on this information, the attack logic in \cref{algo:addattackinfo} $\rightarrow$ \cref{algo:addnodes} starts building the shadow model by adding empty nodes to an empty DT $\mathcal{\tilde{T}}$.  Now, the DT consists of $\operatorname{len}(\mathbf{B})=2$ empty inner nodes and one final leaf node with value $v=0$ (see \cref{init_extraction}). The backlog of incomplete nodes has two entries: $\mathcal{\tilde{T}}.\mathbf{I}=[\mcirc{0}, \mcirc{1}]$.

\begin{algorithm}[t!]
\footnotesize
\caption{DTExtraction}
\label{algo:dtextraction}
\begin{algorithmic}[1]
\Require $\mathbf{R}^l, \mathbf{R}^u, \epsilon, m$
\LineComment{\textcolor{gray}{Run the main attack loop by iteratively crafting inputs, extracting the branching behavior in the TEE and adding the collected information to the shadow DT.}}
\Ensure $\mathbf{\tilde{T}}$%
\State $\beta \gets \bot$ \MyComment{Index to track currently checked feature}
\State $\mathcal{V} \gets \bot$ \MyComment{Node currently targeted for extraction}
\LineComment{1. Step: Explore left-most path}
\State $\mathbf{X}\gets \mathbf{R}^u$
\LineComment{Run inference to get result / attack target DT in TEE to extract branching information}
\State $v,\mathbf{B}\gets \operatorname{AttackPrim}[\mathcal{T}(\bf{X})]$
\State $\mathcal{\tilde{T}}\gets \operatorname{DT}(\bot, [])$ \MyComment{Initialize an empty shadow DT}
\State $\beta\gets\operatorname{AddAttackInfo}(\mathcal{\tilde{T}},\mathcal{V},v,\mathbf{B},\mathbf{X}, m, \beta, \epsilon)$
\LineComment{2. Step: Extract the whole target DT and stop when extraction of the current node is completed and no more incomplete nodes exist}
\While{$!(\mathcal{V}=\bot \wedge \mathcal{\tilde{T}}.\mathbf{I}.\operatorname{empty}())$}
\If{$\mathcal{V}=\bot$}
\State $\mathcal{V}\gets\mathcal{\tilde{T}}.\mathbf{I}.\operatorname{pop}()$
\EndIf
\State $X, \beta\gets \operatorname{CraftNextInput}(\mathcal{V}, \mathcal{\tilde{T}},\mathbf{R}^l,\mathbf{R}^u,\beta, \epsilon)$
\State $v,\mathbf{B}\gets \operatorname{AttackPrim}[\mathcal{T}(\bf{X})]$
\State $\beta\gets\operatorname{AddAttackInfo}(\mathcal{\tilde{T}},\mathcal{V},v,\mathbf{B},\mathbf{X}, m, \beta, \epsilon)$
\EndWhile
\end{algorithmic}
\end{algorithm}

\begin{algorithm}[t!]
\footnotesize
\caption{AddAttackInfo}
\label{algo:addattackinfo}
\begin{algorithmic}[1]
\Require $\mathcal{\tilde{T}}, \mathcal{V}, v, \mathbf{B},\mathbf{X}, m, \beta, \epsilon$
\LineComment{\textcolor{gray}{Add the information collected during the last inference run to the shadow DT. If the feature of the node under attack $\mathcal{V}$ is unknown, extract it from the branching trace or return to trigger another input to the target DT. If the feature is already extracted, determine the node threshold or return to trigger another input for increased threshold resolution.}}
\State $\operatorname{AddNodes}(\mathcal{\tilde{T}}, v, \mathbf{B}, \mathbf{X}, m)$
\If{$\mathcal{V} = \bot$}\MyComment{\textcolor{orange}{Add first path to shadow DT}}
\State \Return $0$
\ElsIf{$\mathcal{V}.f=\bot$}\MyComment{\textcolor{orange}{Determine node feature}}
	\If{$\mathbf{B}.(\mathcal{V}.d) \not = \mathcal{V}.\mathbf{B}.(\mathcal{V}.d)$}\MyComment{Feature detected}
		\State $\mathcal{V}.f\gets \mathbf{S}.(\beta-1)$
		\State \Return $0$
	\Else\MyComment{Feature not detected}
		\State \Return $\beta$
	\EndIf
	\Else\MyComment{\textcolor{orange}{Determine threshold}}
	\State $\delta\gets \mathcal{V}.\mathbf{T}^0.(\mathcal{V}.f)-\mathcal{V}.\mathbf{T}^1.(\mathcal{V}.f)$
	\If{$\delta \leq \epsilon$}
		\State $\mathcal{V}.t\gets\mathcal{V}.\mathbf{T}^0.(\mathcal{V}.f)+(\delta/2)$
		\State $\mathcal{V}\gets\bot$ \MyComment{Set reference null}
	\EndIf
    \State\Return $\beta$
\EndIf
\end{algorithmic}
\end{algorithm}

\begin{algorithm}[t!]
\footnotesize
\caption{AddNodes}
\label{algo:addnodes}
\begin{algorithmic}[1]
\Require $\mathcal{\tilde{T}}, v, \mathbf{B}, \mathbf{X}, m$
\Ensure $\mathcal{\tilde{T}}$
\LineComment{\textcolor{gray}{Traverse the shadow DT based on the branching trace, add non-existing nodes and update threshold ranges.}}
\If{$\mathcal{\tilde{T}}.r=\bot$} \MyComment{\textcolor{orange}{If non-existent, add root node}}
	\State $\mathcal{\tilde{T}}.r\gets \operatorname{Node}(\bot, 0, \mathbf{X}, \mathbf{B})$
	\State $\mathcal{\tilde{T}}.{\mathbf{I}}.\operatorname{append}(\mathcal{\tilde{T}}.r)$
\EndIf
\State $\mathcal{V}\gets \mathcal{\tilde{T}}.r$ \MyComment{Current node}
\For{$i\gets 0; i < m; i\gets i+1$}
	\State $\operatorname{UpdateThresholdRanges}(\mathcal{V}, \mathbf{B}.i, \mathbf{X}, m)$
	\If{$\mathbf{B}_i=0$}
		\If{$\mathcal{V}.l=\bot$}\MyComment{\textcolor{orange}{If non-existent, add left child}}
			\State $\mathcal{V}.l\gets \operatorname{Node}(\mathcal{V}, i+1, \mathbf{X}, \mathbf{B})$
            \State $\mathcal{V}.l.\mathbf{TT}, \mathcal{V}.l.\mathbf{DD}\gets\operatorname{Copy}(\mathcal{V}.\mathbf{TT}), \operatorname{Copy}(\mathcal{V}.\mathbf{DD})$
			\State $\mathcal{\tilde{T}}.\mathbf{I}.\operatorname{append}(\mathcal{V}.l)$
		\EndIf
		\State $\mathcal{V}\gets \mathcal{V}.l$
	\Else
		\If{$\mathcal{V}.r=\bot$} \MyComment{\textcolor{orange}{If non-existent, add right child}}
			\State $\mathcal{V}.r\gets \operatorname{Node}(\mathcal{V}, i+1, \mathbf{X}, \mathbf{B})$
            \State $\mathcal{V}.l.\mathbf{TT}, \mathcal{V}.l.\mathbf{DD}\gets\operatorname{Copy}(\mathcal{V}.\mathbf{TT}), \operatorname{Copy}(\mathcal{V}.\mathbf{DD})$
			\State $\mathcal{\tilde{T}}.\mathbf{I}.\operatorname{append}(\mathcal{V}.r)$
		\EndIf
		\State$\mathcal{V}\gets \mathcal{V}.r$
	\EndIf
\EndFor
\LineComment{Last node is a leaf, remove it from incomplete nodes}
\If{$\mathcal{V}\text{ in } \mathcal{T}.\mathbf{I}$}
	\State $\mathcal{V}.v\gets v$
	\State $\mathcal{T}.\mathbf{I}.\operatorname{remove}(\mathcal{V})$
\EndIf
\end{algorithmic}
\end{algorithm}

\begin{algorithm}[t!]
\footnotesize
\caption{UpdateThresholdRanges}
\label{algo:updatethresholdranges}
\begin{algorithmic}[1]
\Require $\mathcal{V}, b, \mathbf{X}, m$
\LineComment{\textcolor{gray}{Update threshold range of node $\mathcal{V}$ based on input $\mathbf{X}$ and branching information $b$ which indicate wether $\mathcal{V}$ is a left or right child.}}
\If{$\mathcal{V}.{\mathbf{T}^b}.\operatorname{empty}()$} \MyComment{Initialize threshold ranges}
\State $\mathcal{V}.{\mathbf{T}^b}\gets \mathbf{X}$
\Else
\If{$b=0$}\MyComment{$\mathcal{V}$ is left child}
\For{$i\gets 0; i < m; i\gets i+1$}
	\If{$\mathbf{X}.i < \mathcal{V}.{\mathbf{T}^0.i}$}
	\State $\mathcal{V}.{\mathbf{T}^0.i}\gets \mathbf{X}.i$
	\EndIf
\EndFor
\Else\MyComment{$\mathcal{V}$ is right child}
\For{$i\gets 0; i < m; i\gets i+1$}
	\If{$\mathbf{X}.i > \mathcal{V}.{\mathbf{T}^1.i}$}
	\State $\mathcal{V}.{\mathbf{T}^1.i}\gets \mathbf{X}.i$
	\EndIf
\EndFor
\EndIf
\EndIf
\end{algorithmic}
\end{algorithm}

\begin{algorithm}
\footnotesize
\caption{CraftNextInput}
\label{algo:craftnextinput}
\begin{algorithmic}[1]
\Require $\mathcal{V}, \mathcal{\tilde{T}}, \mathbf{R}^l, \mathbf{R}^u, \beta, \epsilon$
\LineComment{\textcolor{gray}{If the feature of the node under attack $\mathcal{V}$ is unknown, create an input to determine it. If the feature is known but the threshold is unknown and $\mathcal{V}$ is not a leaf, create an input to determine the threshold.}}
\If{$\mathcal{V}.f = \bot$}
    \State $\mathbf{\tilde{X}}\gets \operatorname{CraftInpFeature}(\mathcal{V}, \mathcal{\tilde{T}}, \mathbf{R}^u, \mathbf{R}^l, \beta, \epsilon)$
    \State \Return{$\mathbf{\tilde{X}}, \beta+1$}
\EndIf
\If{$\mathcal{V}.t=\bot \wedge \mathcal{V}.v=\bot$}
	\State $\mathbf{\tilde{X}}\gets \operatorname{CraftInpThreshold}(\mathcal{V})$
	\State \Return $\mathbf{\tilde{X}}, \beta$
\EndIf
\end{algorithmic}
\end{algorithm}

\begin{algorithm}[t!]
\footnotesize
\caption{CraftInpFeature}
\label{algo:craftinpfeature}
\begin{algorithmic}[1]
\Require $\mathcal{V}, \mathcal{\tilde{T}}, \mathbf{R}^u, \mathbf{R}^l, \beta, \epsilon$
\LineComment{\textcolor{gray}{Craft next input $\mathbf{\tilde{X}}$ to determine the feature of the node under attack $\mathcal{V}$.}}
\LineComment{\textcolor{orange}{If current node is root, set input for most left path and toggle feature after feature to min. feature value}}
\If{$\mathcal{V} = \mathcal{\tilde{T}}.r$}
	\State $\mathbf{\tilde{X}}\gets \mathbf{R}^u$
	\State $\mathbf{\tilde{X}}_{\beta}\gets \mathbf{R}^l.{\beta}$
\Else\MyComment{If current node is not root}
\State $\mathbf{\tilde{X}}\gets \operatorname{Copy}(\mathcal{V}.\mathbf{X})$
\LineComment{\textcolor{orange}{If there is a node on the path to $\mathcal{V}$ that checks feature $\mathbf{S}.\beta$, craft a new input based on the feature thresholds $\mathcal{V}.\mathbf{DD}$ that reaches $\mathcal{V}$ and toggles its branching decision compared its exploring input $\mathcal{V}.\mathbf{X}$.}}
\If{$\mathcal{V}.\mathbf{TT}.\beta \not = []$}
    \State $tt\gets\mathcal{V}.\mathbf{TT}.\beta$
    \State $dd\gets\mathcal{V}.\mathbf{DD}.\beta$
    \LineComment{Last check on this features went left}
    \If{$\mathcal{V}.\mathbf{B}.(dd.last)=0$}
        \If{$\mathcal{V}.\mathbf{B}.(\mathcal{V}.d)=0$} \MyComment{Current node goes left}
            \State $\mathbf{\tilde{X}}.\beta\gets tt.last+\epsilon$
        \Else
            \LineComment{Find minimal threshold that went right}
            \State $ru \gets \mathbf{R}^u.\beta$
            \State $fb\gets [\mathcal{V}.\mathbf{B}.i \text{ for } i \text{ in } dd]$
            \State $l\gets[t \text{ if } b=1\text{ else } ru \text{ for } (b,t) \text{ in } \operatorname{zip}(fb,tt)]$
            \State $minrt\gets\operatorname{min}(l)$
            \State $\mathbf{\tilde{X}}.\beta\gets minrt - \epsilon$
        \EndIf
    \Else \MyComment{Last check on this features went right}
        \If{$\mathcal{V}.\mathbf{B}.(\mathcal{V}.d)=0$} \MyComment{Current node goes right}
            \State $\mathbf{\tilde{X}}.\beta\gets tt.last-\epsilon$
        \Else
        \LineComment{Find maximal threshold that went left}
        \State $rl\gets \mathbf{R}^l.\beta$
        \State $fb\gets [\mathcal{V}.\mathbf{B}.i \text{ for } i \text{ in } dd]$
        \State $l\gets[t \text{ if } b=0\text{ else } rl \text{ for } (b,t) \text{ in } \operatorname{zip}(fb,tt)]$
        \State $maxlt\gets\operatorname{max}(l)$
        \State $\mathbf{\tilde{X}}.\beta\gets maxlt + \epsilon$
        \EndIf
    \EndIf
\LineComment{\textcolor{orange}{If there is no node on the path to $\mathcal{V}$ that checks feature $\mathbf{S}.\beta$, toggle the branching decision using the limits of the corresponding feature range.}}
\Else
    \If{$\mathbf{B}.\beta = 0$}
        \State $\mathbf{\tilde{X}}.\beta\gets \mathbf{R}^l.\beta$
    \Else
        \State $\mathbf{\tilde{X}}.\beta\gets \mathbf{R}^u.\beta$
    \EndIf
\EndIf
\EndIf
\State \Return{$\mathbf{\tilde{X}}$}

\end{algorithmic}
\end{algorithm}

\begin{algorithm}[t!]
\footnotesize
\caption{CraftInpThreshold}
\label{algo:craftinpthreshold}
\begin{algorithmic}[1]
\Require $\mathcal{V}$
\LineComment{\textcolor{gray}{Craft the next input to determine the threshold of the node under attack $\mathcal{V}$. To narrow down the threshold perform a binary search step based on the current threshold ranges of node $\mathcal{V}$.}}
	\State $\mathbf{\tilde{X}}\gets \operatorname{Copy}(\mathcal{V}.\mathbf{X})$
	\State $\delta\gets \mathcal{V}.\mathbf{T}^0.(\mathcal{V}.f)-\mathcal{V}.\mathbf{T}^1.(\mathcal{V}.f)$
	\State $\mathbf{\tilde{X}}.(\mathcal{V}.f)\gets \mathcal{V}.\mathbf{T}^1.(\mathcal{V}.f) + (\delta/2)$
	\State\Return $\mathbf{\tilde{X}}$
\end{algorithmic}
\end{algorithm}

\subsubsection{Root Node}
Based on the leftmost path already extracted and added to the shadow tree, the attack logic targets the root node $\mathcal{V}=\mcirc{0}$. The extraction process of a node consists of one step to extract the feature and a second step to extract the threshold.

\begin{figure}
    \centering    \includegraphics[width=0.7\linewidth]{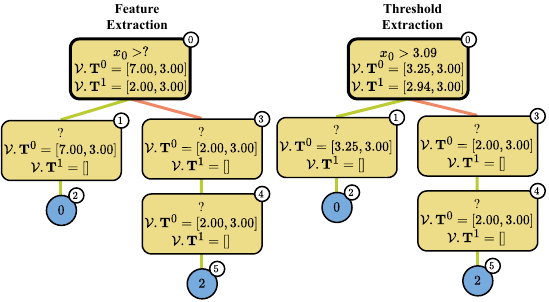}
    \caption{Shadow DT with the extracted root node's feature and threshold.}
    \label{fig:root_node_extraction}
\end{figure}

To extract the feature of node $\mcirc{0}$, the attack logic in \cref{algo:craftnextinput} $\rightarrow$ \cref{algo:craftinpfeature} crafts an input with all feature values set to their upper limits in $\mathbf{R}^u$. Then it toggles one feature - indexed by $\beta$ - after another to their respective minimums in $\mathbf{R}^l$. When the toggling of one feature $s_\beta$ in the input results in a right branch of node $\mcirc{0}$, \cref{algo:addattackinfo} identifies $s_\beta$ as the feature of the node $\mcirc{0}$. Since the root node in our example checks on feature $s_\beta$, with $\beta=0$, the feature extraction requires just a single input $\mathbf{\tilde{X}}=[2.0, 3.0]$ to the target tree and results in a shadow tree as shown on the left side of \cref{fig:root_node_extraction}.

To extract the threshold value of the root node, the attack logic creates inputs based on a \textit{binary search} in \cref{algo:craftnextinput} $\rightarrow$ \cref{algo:craftinpthreshold}. As for our example $\epsilon=0.5$, the threshold search for the root node terminates after four additional inputs: $[4.50, 3.00]$, $[3.25, 3.00]$, $[2.625, 3.00]$ and $[2.938, 3.00]$. The resulting shadow DT is shown on the right side of \cref{fig:root_node_extraction}. During the binary search the attack logic makes use of the threshold ranges $\mathcal{V}.{\mathbf{T}^0}, \mathcal{V}.{\mathbf{T}^1}$ to limit the search space and reducing the number of required queries. Additionally, it performs passive information tracking and updates all threshold ranges along the inference path of all inputs.

\subsubsection{Inner Node}
\begin{figure}[t!]
    \centering    \includegraphics[width=0.7\linewidth]{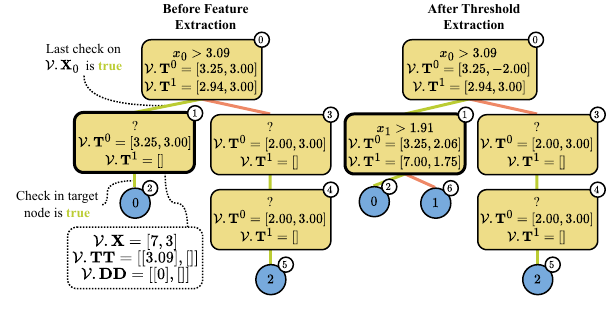}
    \caption{Shadow DT with the extracted inner node's feature and threshold.}
    \label{fig:inner_node_extraction}
\end{figure}

Next, we target the inner node $\mathcal{V}=\mcirc{1}$ shown on the left side in \cref{fig:inner_node_extraction} and start with the extraction of the node feature. For inner nodes, the feature extraction in \cref{algo:craftinpfeature} is more involved. Again, it starts by checking whether $s_0$ is the node feature. First, the logic leverages the feature thresholds to check whether the feature $s_0$ was already tested on the path to the current node: at the index $\beta=0$, the list $\mathcal{V}.\mathbf{TT}$ already contains an entry of length one $[3.094]$ and $\mathcal{V}.\mathbf{DD}$ contains $[0]$ as the root node which already tests feature $s_0$ has depth $0$. Since the input $\mathcal{V}.\mathbf{X}=[7.00, 3.00]$ that initially explored the target node $\mcirc{1}$ traversed left on the last check of feature $\beta$ (in the root node) and also on the check in the target node, the logic generates the input $\mathbf{\tilde{X}}=[3.094+\epsilon, 3.00]$. When the target node tests feature $s_0$, $\mathbf{\tilde{X}}$ would enforce a right traversal confirming the feature to the attacker.

In our example, $\mathbf{\tilde{X}}$ does not result in the inference traversing right after the target node, ruling out $s_0$ as a possible feature. The logic continues generating the next input, checking feature $s_1$. Since on the path to the target node no previous check on $s_1$ is performed, the attack logic toggles the feature to its minimum, resulting in a new input $\mathbf{\tilde{X}}=[7.00, -2.00]$. Performing inference using the new input lets the inference proceeding right after the target node confirming $s_1$ as the feature of the target node. Finally, the threshold is determined using the binary search. The resulting shadow tree is shown on the right in \cref{fig:inner_node_extraction}.

\subsubsection{Duplicated Features} 
\begin{figure}[t!]
    \centering    \includegraphics[width=0.9\linewidth]{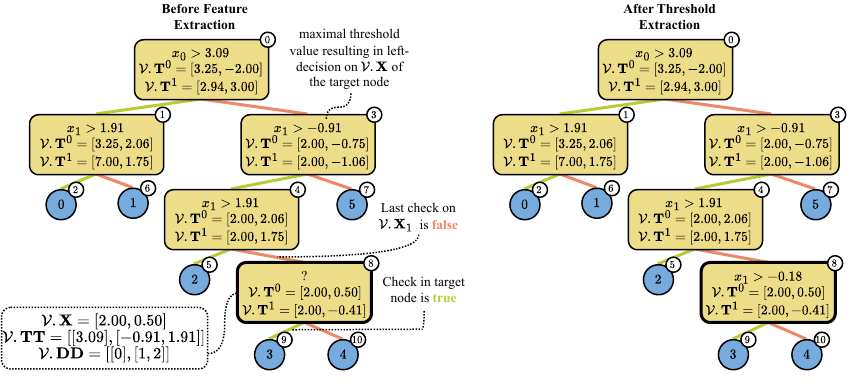}
    \caption{Shadow DT before and after the extraction of the duplicated feature node.}
    \label{fig:duplicated_feature_extraction}
\end{figure}

To describe the extraction of nodes containing features which already occurred along the path from the root node, we start from the shadow tree shown on the left side in \cref{fig:duplicated_feature_extraction}. Now, we extract node $\mathcal{V}=\mcirc{8}$, again starting checking feature $s_0$. 
After ruling out $s_0$, the logic checks whether $s_1$ already occurred earlier on the path from the root node. Given $\mathcal{V}.\mathbf{TT}.{\beta}=[-0.906, 1.906]$, $\mathcal{V}.\mathbf{DD}.{\beta}=[1, 2]$ and $\mathcal{V}.\mathbf{X}=[2.00, 0.50]$ we have the case where the last check of the feature went right and the check on the target node $\mathcal{V}$ went left. In this case, we search for the maximal threshold $maxlt$ checked against $s_1$ that resulted in a left-decision along the path from the root node to the target node.
Setting $\mathbf{\tilde{X}}.1=maxlt+\epsilon=-0.406$ enforces a right-decision and confirms feature $s_1$. The threshold is extracted analog to the above cases.

\subsubsection{Complexity} 
Recall that \cite{wang2025barkbeetle} targets paths that yield distinct predictions (\eg regression trees or classification trees with differentiable leaves) and relies on fault injection, where the required query counts vary hugely with the injection method.
Therefore, we primarily compare \schemename{} with \cite{DBLP:conf/uss/TramerZJRR16}, which is the most closely related work operating under similar extraction assumptions, albeit without considering TEEs.
Let $l$ be the number of leaves in the tree, $d$ the number of continuous features, $h$ the tree depth, and $\epsilon$ the granularity used in binary search. For simplicity, we expect $l = 2^{h-1}$.
For continuous features within the range $[0, b]$, determining a single threshold requires at most $\log_2{\frac{b}{\epsilon}}$ queries. 
The complexity of \cite{DBLP:conf/uss/TramerZJRR16} is $O(l^2 \cdot d \cdot \log_2{\frac{b}{\epsilon}})$.
\schemename{}, requires $d$ queries per inner node to detect the feature and a binary search to find the corresponding threshold, resulting in a complexity of $O(d\cdot l+l\cdot\log_2{\frac{b}{\epsilon}})$.
Overall, in contrast to \cite{DBLP:conf/uss/TramerZJRR16}, which performs isolated binary searches for each feature range, \schemename{} utilizes side-channel branching traces to continuously prune the search space. This difference enables \schemename{} to extract the tree with significantly fewer queries.

\subsection{Attack Primitives}
\subsubsection{SGX}

To attack DT inference in Intel SGX enclaves (release, not debug enclaves) on current Intel Xeon server CPUs, we first extract the CBP's PHR. The PHR is an interesting target as doublets caused by enclave execution remain there even after the enclave exit~\cite{yavarzadeh2024pathfinder}. Based on the PHR's  content, we reconstruct the execution path through the tree.

\noindent\textbf{PHR Extraction.} Based on previous work, we can read the content of the PHR by using a \texttt{READ\_PHR} helper ~\cite{mahling2023reverse,yavarzadeh2024pathfinder,yavarzadeh2023half}. We utilize the \texttt{WRITE\_PHR} (writes arbitrary content into the PHR), \texttt{SHIFT\_PHR} (shifts the PHR by a specified number of doublets), and \texttt{CLEAR\_PHR} (clears the PHR by zeroing all doublets) helpers from previous work~\cite{yavarzadeh2024pathfinder}. Before employing these helpers on current Intel Xeon 6 server hardware, we verified that the CBP on newer Intel processors matches the behavior on the latest reverse-engineered generation, Raptor Lake. By implementing experiments from previous work~\cite{yavarzadeh2023half, mahling2023reverse}, we observed no significant changes.

\begin{figure}
    \centering
    \includegraphics[width=0.6\linewidth]{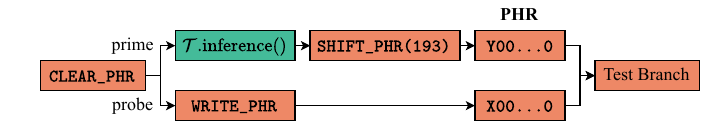}
    \caption{Illustration of the \texttt{READ\_PHR} helper while extracting the first doublet of the PHR.}
    \label{fig:read_phr}
\end{figure}

We extract the PHR content doublet by doublet through systematically enforced collisions in the PHT (see also \cref{fig:read_phr}). After clearing the PHR, we alternately execute a prime and a probe path. On the prime path, we execute the victim code, the DT inference in our case. To extract the first doublet, we shift the PHR by 193 positions, isolating the previously first doublet in the last position while all other doublets are shifted out. Hence, the rest of the PHR is in a known state containing only zero doublets. The only unknown value is the isolated doublet $Y$, which has four possible values. In the end we execute a test branch such that it results in a non-taken outcome.%

On the probe path, we test each possible $X \in \{0,1,2,3\}$. For each test, we write $[X,0,...,0]$ into the PHR and execute the test branch such that it results in a taken outcome. Since we use the same test branch on both paths, the PHR is the only distinguishing factor for the PHT index. Two outcomes are possible: if $X$ equals $Y$ from the prime path, both paths generate identical PHRs and therefore identical PHT indices. However, since the test branch outcomes are opposite (taken vs. not taken), they collide in the same PHT entry. Performing this procedure multiple times allows observing a measurable spike in miss-predictions via hardware performance counters. Conversely, if $X\not = Y$, the PHRs differ too, resulting in different PHT entries and the miss-predictions remain low, as there is no collision in the PHT.

To extract subsequent doublets, we decrement the shift amount iteratively. For the second doublet, we shift by 192, isolating both the first and second doublet at the end of the PHR. Since we already extracted the first doublet, only the second remains unknown. We can repeat the previously described procedure to extract this unknown doublet as well. That way, we can recover the full PHR. The \texttt{READ\_PHR} helper is inspired by the primitive from Yavarzadeh et al.~\cite{yavarzadeh2024pathfinder}. However, we simplified the PHR read by removing the training branch used in previous work \cite{yavarzadeh2024pathfinder}. %

Given all possible execution paths and leveraging a \texttt{READ\_PHR} helper Yavarzadeh et al.~\cite{yavarzadeh2024pathfinder} demonstrate the successful recovery of an image from the libjpeg library. In our setting we do not know all possible execution paths through the targeted DT libraries, as this information directly corresponds to the model architecture. However, in the following we show that this information is not required to successfully steal DTs.

\noindent\textbf{Reconstruction of the DT's Execution Path.} Given the extracted PHR, we reconstruct the execution path through the DT. A node traversal is typically implemented as an \texttt{if/else} statement, which compilers often realize using a single conditional branch. When the \texttt{if}-condition evaluates to false, execution proceeds to the \texttt{else} block by taking the branch. When the condition evaluates to true, the branch is not taken and execution continues with the \texttt{then} block immediately following the branch instruction. Since the \texttt{else} block is placed after the \texttt{then} block, the compiler inserts an unconditional jump at the end of the \texttt{then} block to skip the \texttt{else} block.

This behavior produces distinct PHR patterns for each execution path. In the taken case (\ie the \texttt{else} block), the conditional branch itself updates the PHR. In the not-taken case (\ie the \texttt{then} block), the conditional branch leaves the PHR unchanged, but the subsequent unconditional follow-up jump updates it. Therefore, each outcome of a compiled \texttt{if/else} statement generates a unique PHR pattern. Section~\ref{sec:vulnerable_libraries} demonstrates that libraries such as mlpack and OpenCV exhibit these patterns, enabling DT extraction via the PHR.

\subsubsection{AMD SEV}
Gast et al. show that performance counters can be used in combination with single-stepping to retrieve branch-dependent code execution~\cite{DBLP:conf/ndss/GastWSG25} using the SEV-Step framework. We adapt the same technique to observe the branching behavior of our target DTs. For each single-step event, we add the configuration of performance counters to monitor the performance counter events \textit{Retired Taken Branch Instructions} and \textit{Retired Conditional Branch Instructions}. We assume that the attacker uses the established technique of monitoring a page-based controlled-channel, either by manipulating PTEs~\cite{DBLP:conf/sp/XuCP15} and/or observing page-table-walks and monitoring the ``accessed'' and ``dirty'' bits of a page~\cite{van2017telling, DBLP:conf/uss/MoghimiBHPS20}, to locate all relevant code pages and their invocation order. This serves as a signal for enabling single-stepping on the target code page such that the attacker can collect the vulnerable branch.

\subsection{Vulnerable Libraries}
\label{sec:vulnerable_libraries}
We reveal vulnerable branch-dependent behavior in the following libraries (see also \cref{tab:jcc-comparison}).
To compile the libraries, we use Clang 12.0.1 and GCC 11.4.0.

\noindent\textbf{OpenCV.}
We evaluate OpenCV~\cite{bradski2000opencv} (84.5k GitHub stars) version 4.12.0 using common CMake targets: \texttt{Release} leverages optimization level O3 and \texttt{Debug}  O0.
The vulnerable branch is present in $\texttt{predictTrees()}$\footnote{\href{https://github.com/opencv/opencv/blob/49486f61fb25722cbcf586b7f4320921d46fb38e/modules/ml/src/tree.cpp\#L1433}{GitHub permalink (accessed Oct~17,~2025)}}, which emits the vulnerable JCC/\texttt{jmp} sequence when using the Debug preset. 
This two-step pattern is required for the SGX attack and cannot be observed in any of the release setups we tested. Notably, the GCC/Release setup is exploitable on SEV.

\noindent\textbf{mlpack.} We evaluate the mlpack \cite{DBLP:journals/jossw/CurtinESABBBDEGJKKMSSPSS23} (5.5k GitHub stars), using stable release 4.6.2. The vulnerable branching code can be found in the $\texttt{ChooseDirection()}$\footnote{\href{https://github.com/mlpack/mlpack/blob/0fdccbfb21e142f873c8d710c12ff2ebf3c9dbfb/src/mlpack/methods/decision_tree/splits/best_binary_numeric_split_impl.hpp\#L533}{GitHub permalink (accessed Oct~17,~2025)}} functionality. 
Here, when compiled with optimization level O0, a conditional jump is followed by a fall-through \texttt{jmp} instruction. 

\noindent\textbf{emlearn.} We evaluate emlearn \cite{emlearn} (673 GitHub stars) release 0.21.1. Similar to previous work \cite{wang2025barkbeetle} we compile the inference code generated by emlearn with optimization level O0. Higher optimization level might also be vulnerable, but for brevity we skip the analysis and focus on the more popular libraries.

\begin{table}
\centering
\footnotesize
\caption{Jump-condition outcomes across compilers and optimization levels. Red marked outcomes are vulnerable. If both columns are red, the library is vulnerable in Intel SGX, otherwise its only vulnerable in AMD SEV.}
\label{tab:jcc-comparison}

\setlength{\tabcolsep}{4pt}%

\begin{tabular}{@{}llcc@{\hskip 10pt}llcc@{}}
\toprule
\multicolumn{4}{@{}c@{\hskip 10pt}}{\textit{(a) mlpack}} &
\multicolumn{4}{c@{}}{\textit{(b) OpenCV}} \\
\cmidrule(r){1-4} \cmidrule(l){5-8}
\textbf{Comp.} & \textbf{Opt.} & \textbf{JCC} & \textbf{Jmp} &
\textbf{Comp.} & \textbf{Opt.} & \textbf{JCC} & \textbf{Jmp} \\
\midrule
GCC   & \texttt{O0}         & \cellcolor{orange}\texttt{je}  & \cellcolor{orange}Yes &
Clang & \texttt{Debug}      & \cellcolor{orange}\texttt{jb}  & \cellcolor{orange}Yes \\
GCC   & \texttt{O1--O3, Os} & \cellcolor{mint}None           & \cellcolor{mint}No  &
GCC   & \texttt{Debug}      & \cellcolor{orange}\texttt{jb}  & \cellcolor{orange}Yes \\
Clang & \texttt{O0}         & \cellcolor{orange}\texttt{jb}  & \cellcolor{orange}Yes &
Clang & \texttt{Release}    & \cellcolor{mint}None           & \cellcolor{mint}No  \\
Clang & \texttt{O1}         & \cellcolor{orange}\texttt{jae} & \cellcolor{mint}No  &
GCC   & \texttt{Release}    & \cellcolor{orange}\texttt{jb}  & \cellcolor{mint}No  \\
Clang & \texttt{O2, O3, Os} & \cellcolor{orange}\texttt{jae} & \cellcolor{orange}Yes &
      &                     &                                &                      \\
\bottomrule
\end{tabular}
\end{table}

\section{Evaluation}
In this section we first present our evaluation of \schemename's attack logic and then compose the attack primitives with the attack logic to demonstrate end-to-end attacks. We used public available datasets from the UCI ML Repository and Kaggle\footnote{\href{https://archive.ics.uci.edu/dataset/53/iris}{iris}, \href{https://archive.ics.uci.edu/dataset/374/appliances+energy+prediction}{appliances}, \href{https://archive.ics.uci.edu/dataset/94/spambase}{spam},
\href{https://archive.ics.uci.edu/dataset/17/breast+cancer+wisconsin+diagnostic}{breast cancer},
\href{https://archive.ics.uci.edu/dataset/206/relative+location+of+ct+slices+on+axial+axis}{CT Slices},
\href{https://archive.ics.uci.edu/dataset/264/eeg+eye+state}{EEG Eye},
\href{https://archive.ics.uci.edu/dataset/174/parkinsons}{parkinsons},
\href{https://archive.ics.uci.edu/dataset/96/spectf+heart}{SPECTF Heart},
\href{https://archive.ics.uci.edu/dataset/75/musk+version+2}{ Musk v2},
\href{https://www.kaggle.com/datasets/mathchi/diabetes-data-set?resource=download}{diabetes}
}.

\subsection{Attack Logic}

We implemented \schemename's attack logic as stand-alone Python package. In the following, we evaluate \schemename's extraction fidelity similar to previous work by comparing the predictions of the target DT $\mathcal{T}$ with the predictions of the shadow model $\mathcal{\tilde{T}}$. We use the training data distribution $D$ and compute the extraction error as
$R(\mathcal{T}, \mathcal{\tilde{T}})=\sum_{(\mathbf(X),y)\in D}d(\mathcal{T}(\mathbf{X}),\mathcal{\tilde{T}}(\mathbf{X}))/|D|$.
Here $d$ denotes the 0-1 measure and we compute $d(y, \tilde{y})=1$ if $y=\tilde{y}$ and otherwise $d(y,\tilde{y})=0$. The extraction accuracy is defined as $1-R$.

\begin{table*}[htbp]
\centering
\footnotesize
\caption{Performance comparison of \schemename{} and \textsc{APIAttack} \cite{DBLP:conf/uss/TramerZJRR16} for different datasets and DT models. We report the number of features (used in the DT) and classes ($\mathbb{R}$ indicates a regression task) per dataset as well as the number of nodes, leaves and the depth of the trained models. As performance measure, we use the best data point from the Pareto frontier (see \cref{fig:pareto_frontier_cost_accuracy}) that maximizes the model fidelity.}
\label{tab:dt_performance_comparison}
\setlength{\tabcolsep}{2.7pt}
\begin{adjustbox}{width=\textwidth}
\begin{tabular}{llllllrr}
\toprule
\multirow{2}{*}{\textbf{Model}} 
  & \multirow{2}{*}{\textbf{\# Features}} 
  & \multirow{2}{*}{\textbf{\# Classes}}
  & \multirow{2}{*}{\textbf{\# Nodes}}
  & \multirow{2}{*}{\textbf{\# Leaves}} 
  & \multirow{2}{*}{\textbf{Depth}}
  & \multicolumn{2}{c}{\textbf{\# Queries / $1-R$}} \\[0.5ex]
\cmidrule(lr){7-8}
 &  &  &  &  &
  & \textsc{APIAttack}
  & \schemename \\ 
\midrule
    Appliances & 25 & $\mathbb{R}$ & 315 & 158 & 17 & 28,088 / 1.00 & \textbf{4,090 / 1.00} \\
    CT Slices & 144 & $\mathbb{R}$ & 469 & 235 & 32 & 145 / 0.03 & \textbf{20,632 / 1.00} \\
    Musk V2 & 70 & 2 & 231 & 116 & 19 & 9,386 / 0.84 & \textbf{9,858 / 1.00} \\
    Diabetes & 8 & 2 & 251 & 126 & 17 & 193 / 0.54 & \textbf{6,508 / 1.00} \\
    Iris & 3 & 3 & 17 & 9 & 5 & 24 / 0.95 & \textbf{55 / 1.00} \\
    Spam & 45 & 2 & 413 & 207 & 29 & 3,305 / 0.82 & \textbf{12,165 / 1.00} \\
    Breast Cancer & 13 & 2 & 39 & 20 & 7 & 597 / 0.81 & \textbf{305 / 1.00} \\
    SPECTF Heart & 21 & 2 & 51 & 26 & 10 & 438 / 0.73 & \textbf{1,534 / 1.00} \\
    Parkinsons & 10 & 2 & 33 & 17 & 8 & 407 / 0.89 & \textbf{296 / 1.00} \\
    EEG Eye & 14 & 2 & 453 & 227 & 19 & 3,612 / 0.64 & \textbf{5,896 / 1.00} \\
\bottomrule
\end{tabular}
\end{adjustbox}
\end{table*}

To contextualize the evaluation results of \schemename{} we compare them to the black-box attack of \etal{Tramèr} \cite{DBLP:conf/uss/TramerZJRR16}, which we note as \textsc{APIAttack}.
To evaluate on realistic configurations and DT models and similar to \etal{Tramèr}, we trained all evaluated models using BigML\footnote{\href{https://bigml.com}{https://bigml.com}} and the platform default settings. The structural information of the trained DTs, including node counts, leaves, and depths, are detailed in \cref{tab:dt_performance_comparison}.

We designed the following experiment: for both methods we set an initial extraction resolution of $\epsilon=100$. Then we perform the attack and measure the extraction accuracy and the number of performed queries. After the first run we half $\epsilon$ and run the attack again. We repeat until an attack run achieves an extraction accuracy of 100\%, the run takes longer than one hour, or 10 consecutive runs result in the same extraction accuracy. For the \textsc{APIAttack} we use the open available proof of concept implementation.

\begin{figure*}
    \centering    \includegraphics[width=\linewidth]{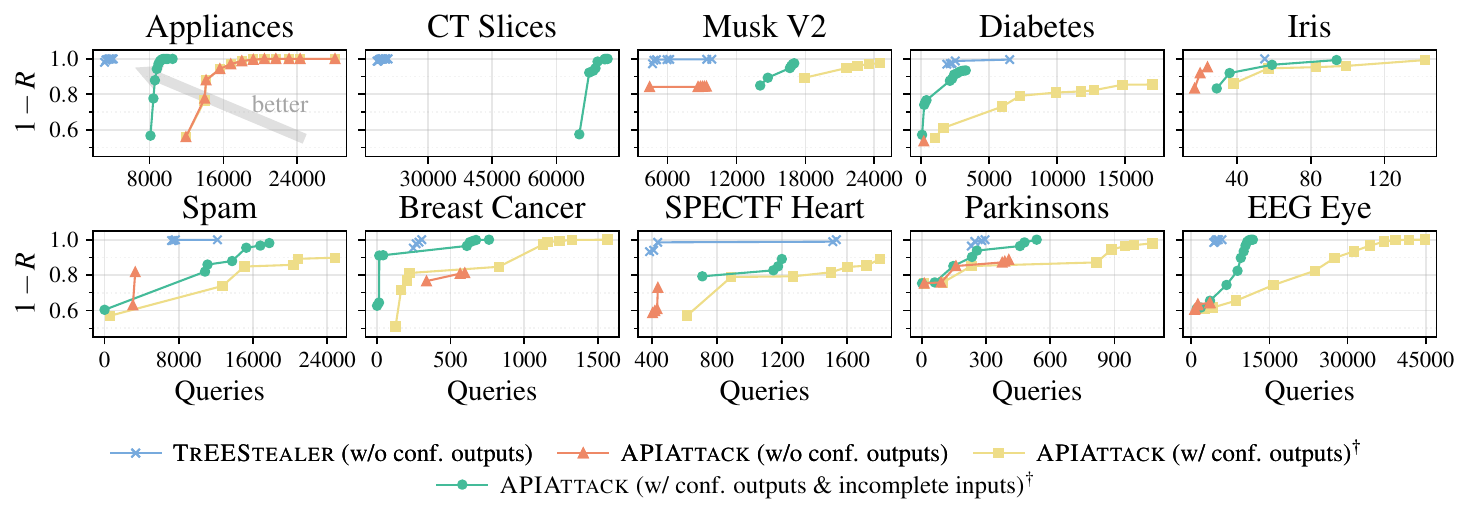}
    \caption{Pareto frontier of the cost vs. extraction accuracy tradeoff. Note, that only data points with $1-R > 0.5$ are shown and attacks marked with $\dagger$ expect a stronger attacker compared to \schemename.}
    \label{fig:pareto_frontier_cost_accuracy}
\end{figure*}

\cref{fig:pareto_frontier_cost_accuracy} shows the Pareto frontiers of the cost accuracy trade-off behind the model stealing attacks derived via the above experiment. We note that \schemename{} achieves a higher extraction accuracy for all evaluated DT models except for the appliances dataset. 
\cref{tab:dt_performance_comparison} shows that the extraction accuracy of \schemename{} is always perfect. Besides, the Pareto frontier of \schemename{} always significantly exceeds the one of \textsc{APIAttack}.

One would expect the black-box \textsc{APIAttack} to perform best on regression DTs, as in that case all outputs are unique. As shown in \cref{tab:dt_performance_comparison}, \textsc{APIAttack} achieves perfect extraction accuracy for the appliances dataset, but fails to extract the CT Slices DT because its excessive query generation violates the timeout of the experiment.  As of writing, larger regression DTs require a side-channel, like the one exploited by \schemename, for effective extraction. In comparison to regression DTs, \textsc{APIAttack} extracts larger classification DTs more effectively, but \schemename{} still achieves better or equal extraction accuracy.

In the classical extraction setting described in our threat model, the attacker only receives the plain inference outputs from the target model. Beyond that, the \textsc{APIAttack} can additionally leverage rich API outputs to boost the attack performance. Although such a setting equates to a much weaker threat model, as such information is most times neither required nor provided, we add corresponding results to our evaluation. \cref{fig:pareto_frontier_cost_accuracy} shows that additional confidence values in the API output can significantly improve the extraction accuracy and efficiency. APIs that additionally enable to submit incomplete queries and output terminating inner nodes allow for even more effective extraction. While \schemename{} is still more efficient (without such rich API outputs) and often also more effective, an attacker might prefer \textsc{APIAttack} when such rich API information is available and side channels are expensive to acquire.

\subsection{End-to-End Enclave Attacks}

\subsubsection{SGX} 
\noindent\textbf{Setup.} For SGX experiments we used an Intel(R) Xeon(R) 6520P (Granite Rapids). The host runs Ubuntu 24.04.3 LTS with the 6.8.0-87-generic Linux kernel version.

\definecolor{myorange}{RGB}{255,165,0}
\definecolor{myturquoise}{RGB}{64,224,208}
\definecolor{mypurple}{RGB}{128,0,128}
\noindent\textbf{mlpack.} The \texttt{CalculateDirection} function employs a conditional branch to determine left or right traversal at each node. As discussed in Section~\ref{sec:vulnerable_libraries}, compilers generate a follow-up jump instruction that creates distinguishable PHR patterns for left and right traversal paths. mlpack invokes \texttt{CalculateDirection} recursively for each node until reaching a leaf. Each iteration produces nine branch updates to the PHR: eight branches execute consistently and generate identical PHR patterns due to their fixed branch and target addresses. The ninth branch differs based on the traversal direction. It is either the conditional branch when taken (right traversal) or the follow-up jump when not taken (left traversal). We can identify distinguishable 9-doublet patterns in the PHR for a left/right traversal. Table~\ref{tab:phr_sequences_example} illustrates the PHR value of two classification DT traversals in mlpack.

Based on these patterns, we can extract DTs with depths up to eleven nodes. The limitation stems from the PHR's fixed capacity of 194 doublets. The SGX and mlpack exit procedures consume the 103 most recent doublets in the PHR, leaving only 91 doublets that reflect the actual node traversal. This remaining space fits up to ten 9-doublet patterns and a final single doublet for the branching decision after the root node. While left for future work, applying the extended \texttt{READ\_PHR} helper of Yavarzadeh et al.~\cite{yavarzadeh2024pathfinder} to \schemename{} would enable the extraction of DTs with even more than eleven nodes.

\newcommand{\tinycolorbox}[2]{\tikz[baseline=(a.base)]\node[fill=#1, inner xsep=0pt](a){#2};}

\begin{table}[t]
  \centering
  \footnotesize
  \begin{tabular}{ll}
    \toprule
    DT Execution Path & Relevant PHR Part \\
    \midrule
    $\rightarrow\,$\texttt{\tinycolorbox{mint}{L}\tinycolorbox{pear}{LLLL}} & \texttt{\tinycolorbox{pear}{3}03101302 \tinycolorbox{pear}{3}03101302 \tinycolorbox{pear}{3}03101302 \tinycolorbox{pear}{3}03101302 \tinycolorbox{mint}{3}}$\,\leftarrow$ \\
    $\rightarrow\,$\texttt{\tinycolorbox{pink}{R}\tinycolorbox{pear}{L}\tinycolorbox{orange}{R}\tinycolorbox{pear}{L}\tinycolorbox{orange}{R}} & \texttt{\tinycolorbox{orange}{2}03101302 \tinycolorbox{pear}{3}03101302 \tinycolorbox{orange}{2}03101302 \tinycolorbox{pear}{3}03101302 \tinycolorbox{pink}{2}}$\,\leftarrow$ \\
    \bottomrule
  \end{tabular}
  \caption{Illustration of execution paths through an mlpack classification DT and the extracted PHR values. The root node is indicated with an arrow.}
  \label{tab:phr_sequences_example}
\end{table}

\subsubsection{SEV}
\noindent\textbf{Setup.} We conduct experiments on a server with an AMD EPYC 7763 64-core processor, running Ubuntu 22.04.4 LTS with a custom kernel (SEV-Step 5.19.0-rc6). The SEV-SNP VM executes Ubuntu 22.04.2 LTS with kernel 6.8.0, hosting a HTTP server that processes inference inputs from a host-based client via POST requests. CPU cores are isolated and frequency fixed. We test statically linked mlpack and dynamically linked emlearn/openCV by sending inputs via HTTP while single-stepping the VM.

\noindent\textbf{emlearn.} The target binary's main function calls a tree implementation via a shared library. First, we track the main function's page. When this is hit, we begin tracking the tree page. Upon hitting the tree page, we enable single-stepping and track all other pages. For multi-page trees, single-stepping continues across consecutive tree code pages. For each single-step, we use the \textit{Retired Taken Branch} and \textit{Retired Conditional Branch} performance counter events to determine tree branch outcomes.

\noindent\textbf{mlpack.} We identify a sequence of 4 pages to the target page $x$ with \texttt{CalculateDirection}. This is called by the previous page $x-1$ which loops over the inputs given. Hence, we disable single-stepping at $x-1$, while tracking $x$ and $x-2$. When $x$ is hit, we enable single-stepping, while tracking $x-1$. Tracing is terminated when $x-1$ returns to $x-2$. The vulnerable jump is identifiable in traces due to occurring at a fixed offset and its branching behavior is evaluated using the performance counters.

\noindent\textbf{OpenCV.} We identify three function calls over two pages to our target function. To differentiate execution in the target function from its wrapper, which is co-located on the same page, we use two different page tracking sequences. OpenCV classifies all inputs in the target function over a loop. To identify the vulnerable jump in our trace, we search at fixed offsets for conditional jumps, while also using the performance counter events to identify absolute jumps that indicate a loop. %

\section{Countermeasures}

A straightforward defense against model extraction is to limit anomaly MLaaS API predictions.
In practice, however, the MLaaS API must remain usable and widely accessible for large volumes of queries. 
Suppressing confidence scores can effectively prevent attacks such as \cite{DBLP:conf/uss/TramerZJRR16}, but this is ineffective against \schemename{}.
Differential privacy (DP) primarily protects training data rather than the model parameters, and therefore does not prevent the extraction of tree structure or node thresholds. 
In this section, we discuss potential countermeasures against TEE-based attack primitives and ensemble methods.

\subsection{Data-oblivious ML Algorithms}
Recall that \schemename{} relies on TEE-based primitives to observe branching behavior during DT inference. 
One promising mitigation is to eliminate data-dependent control flow in the inference program. 
Prior work on data-oblivious ML for Intel SGX~\cite{ohrimenko2016oblivious,law2020secure,wang2022enclavetree} follows this approach by redesigning program so that execution does not depend on secret inputs. 
These systems rely on oblivious primitives implemented in x86-64 assembly, where all operations are performed on registers and all memory accesses follow deterministic patterns. 
Since registers are private to the processor, register-to-register computation is data-oblivious by default.
For example, \cite{law2020secure,wang2022enclavetree} redesign XGBoost and Hoeffding Tree algorithms such that every decision node is evaluated using constant-time and data-independent logic, removing observable branching inside SGX. 
Incorporating similar data-oblivious primitives into DT inference would prevent \schemename{} from learning branch outcomes, thereby limiting or stopping tree reconstruction.

\subsection{Countermeasures against Microarchitectural Attacks for TEEs}
Regarding countermeasures against single-stepping-based attack primitives~\cite{DBLP:conf/ndss/GastWSG25,DBLP:journals/tches/WilkeWRE24}, a straightforward mitigation is to restrict or disable hardware performance counters, since these counters enable precise instruction-level tracing when combined with frequent interrupts. 
AMD’s proposed Performance Monitoring Counter Virtualization may limit hypervisor visibility into guest counter values, but this feature is not yet broadly deployed~\cite{amd2024manual}. 
SEV-SNP also provides SecureTSC, which allows a VM to detect unusually frequent interrupts indicative of single-stepping. 

Mitigating PHR/PHT-based leakage requires either preventing predictor state from being shared across enclave boundaries or eliminating data-dependent control flow inside the enclave. Hardware-level defenses such as branch predictor isolation or state flushing on enclave transitions~\cite{vanbulck2018foreshadow,brasser2017sanctum} can reduce predictor-based leakage, but require architectural support and are not available in current SGX deployments. Software-level approaches, as discussed previously, replace conditional branches with data-oblivious primitives to avoid secret-dependent predictor training~\cite{ohrimenko2016oblivious,law2020secure,wang2022enclavetree}. Enclaves may also leverage trusted timing sources to detect unusually frequent interrupts indicative of single-stepping~\cite{DBLP:conf/ndss/GastWSG25}, though this only provides detection rather than prevention. Overall, mitigating PHR/PHT leakage in SGX requires either data-oblivious execution or hardware changes, both introducing non-trivial performance overheads.

\subsection{Ensemble Methods}
Prior work \cite{DBLP:conf/uss/TramerZJRR16,wang2025barkbeetle} often rely on the model's returned prediction and thus cannot recover individual trees within an ensemble, since the final prediction is aggregated (majority vote for classification, or mean for regression).
For this reason, ensemble methods are commonly viewed as a practical mitigation.
In contrast, \schemename{} leverages branching behavior of each internal node. From these traces we can recover: (i) which features are used in an individual tree (\ie bootstrap sampling and random feature selection); and (ii) the tree topology, node ordering, and numeric split thresholds by adjusting inputs.
Labels of each path remain hidden by the ensemble aggregation.
Nevertheless, labels can be estimated probabilistically: we first assign a candidate label to each reconstructed path, then iteratively refine these assignments by comparing the recovered forest’s predictions to the target model.
The same approach applies to regression forests, although the estimation accuracy may degrade due to continuous output averaging.
We expect this procedure to recover leaf labels with low error in practice while a full analysis of accuracy and convergence is left to future work.

\section{Related Work}
We introduce microarchitectural attacks targeting TEEs and summarize related model extraction attacks.

\subsection{Microarchitectural Attack Primitives for TEEs}
Fault-less page table attacks~\cite{kim2019sgx,van2017telling,wang2017leaky} exploit page tables directly rather than relying on page faults. By inspecting page-table attributes or the caching behavior of unprotected page-table memory, an attacker can infer enclave memory-access patterns, although only at page granularity. Segmentation-based attacks~\cite{gyselinck2018off} obtain finer resolution by manipulating the segmentation unit, but only apply to 32-bit enclaves, as segmentation is disabled in 64-bit mode~\cite{fei2021security}.
Branch-prediction structures also provide attack surface in SGX. Residual predictor state from the LBR and BTB enables fine-grained control-flow reconstruction. Branch shadowing~\cite{lee2017inferring} uses timing differences and predictor collisions to reconstruct an enclave’s control-flow graph and recover cryptographic secrets; the same work showed that combining LBR leakage with APIC-based interrupt control enables reliable single-stepping RSA key extraction.

\subsection{Model Extraction Attacks}
Since the pioneering work \cite{DBLP:conf/uss/TramerZJRR16}, most research has focused on query-based approaches, where the adversary interacts solely with black-box APIs.
Beyond this, a growing body of work investigates how side-channel leakages from the underlying ML system implementations can be exploited rather than solely querying the model interface.
In this section, we mainly discuss this line of research, as it is most closely related to \schemename.

\subsubsection{Query-based Attacks}

Earlier research has primarily focused on NN models~\cite{wu2022model,jagielski2020high,liang2024model,chandrasekaran2020exploring} with a primary goal of improving the accuracy of the stolen model and reducing query complexity.
Only a few works~\cite{DBLP:conf/uss/TramerZJRR16,chandrasekaran2020exploring,oksuz2024autolycus} have studied extraction attacks on DT models.
In \textsc{APIAttack}~\cite{DBLP:conf/uss/TramerZJRR16}, 
the attacker records leaf identifiers and infers the decision boundaries required to remain in that leaf. 
By iteratively generating queries that explore unexplored leaves, the adversary can recover the full tree structure. 
The query complexity can be further reduced by leveraging the ``incomplete query", where selected input features are left unspecified. 
This optimization, however, can be trivially prevented by disabling support for incomplete queries.  
Subsequent work by \cite{chandrasekaran2020exploring} relaxes these assumptions, but requires even more queries and achieves just a slight improvement in extraction accuracy compared to \textsc{APIAttack}. 
More recently, \cite{oksuz2024autolycus} leverages explainable AI techniques to extract interpretable models such as DT. 
By exploiting explanations of decision boundaries, they construct surrogate models that approximate the target. 
Their experiments report substantially greater efficiency than \cite{DBLP:conf/uss/TramerZJRR16,chandrasekaran2020exploring}. 
Nevertheless, the practicality of this method is constrained by its reliance on rich auxiliary information from the MLaaS platform, which may be restricted or directly targeted as part of the adversary’s objectives in DT extraction.

\subsubsection{Side-Channel-based Attacks}
Relying only on query-response pairs limits the effectiveness of extraction attacks.
For example, regarding NN models with many layers, it is non-trivial to extract parameters in intermediate layers using only observable predictions. 
Recent studies instead exploit side-channel leakages observed during inference, such as memory-hierarchy patterns \cite{yan2020cache,liu2020ganred,hu2020deepsniffer,liu2024deepcache}, power consumption patterns \cite{gao2024deeptheft,dubey2022high}, electromagnetic emanations \cite{batina2019csi,yu2020deepem}, or even direct interference with hardware such as fault injection \cite{breier2022sniff,rakin2022deepsteal,hector2023fault} where the attacker exploits bit flips in certain layers.
\cite{DBLP:journals/corr/abs-2503-19142} studies page access patterns triggered by non-constant-time activation functions (e.g., $\mathtt{exp()}$) inside SGX using SGX-Step \cite{DBLP:conf/sosp/BulckPS17}. 
The attacker then conducts a binary search over controlled neuron inputs to search for the scaled value.
This work is related to \schemename, which also builds on stepping-based primitives to recover model parameters, but focuses mainly on extracting first-layer weights and biases of NN models in the TensorFlow Microlite library   and does not consider tree-based models.

More recently, BarkBeetle \cite{wang2025barkbeetle} extracts DT models using fault injection, combined with access to the query-response pairs.
Compared to query-based extraction approaches \cite{DBLP:conf/uss/TramerZJRR16,chandrasekaran2020exploring,oksuz2024autolycus}, BarkBeetle reduces the number of queries and does not require rich auxiliary information from the MLaaS platforms. 
In addition to decision boundaries, it also recovers feature importance along each path (i.e., repeated feature usage). 
Its extraction goal is similar to \schemename, but the query overhead depends heavily on the used fault injection technique. 
For example, in their practical evaluation, voltage glitching requires several hundred extra queries. 
In contrast, \schemename ~achieves more stable and scalable extraction in TEE-protected scenarios.

\section{Conclusion}
In this paper, we introduce \schemename{}, which is, to the best of our knowledge, the first high-fidelity DT extraction attack using microarchitectural leakage. \schemename{} can reconstruct DTs without the need for rich outputs, by observing side-channel information from TEE execution alongside minimal API queries. We demonstrate this on commercially available TEEs, namely SGX and SEV. To conclude, we demonstrate that TEEs do not prevent DT extraction, undermining their promise of model confidentiality.

\section{Acknowledgments}
This work has been supported by the BMFTR through the project AnoMed-II and by the EPSRC under grant EP/X03738X/1.

\appendix

\bibliographystyle{alpha}
\bibliography{bib.bib}

\end{document}

%% file: figs/treestealer_diagram.tex
\begin{tikzpicture}[
    font=\scriptsize,
    box/.style={
        draw,
        node distance=3px,
        minimum width=80pt,
        minimum height=40pt,
    },
    attacker/.style={
        box,
        fill=orange,
    },
    victim/.style={
        box,
        fill=mint,
    },
    arrow/.style = {thick, ->, >=stealth},
    treenode/.style={circle, draw, fill=black, minimum size=1mm, inner sep=0pt},
    leafnode/.style={},
    level distance=.35cm,
    level 1/.style={sibling distance=.35cm},
    level 2/.style={sibling distance=.2cm},
]
    
    \node[]                                     (title1) {\textbf{Attack Primitive}};
    \node[attacker, below=.1cm of title1.south, minimum height=20pt] (att) {\textbf{SEV-Step} or \textbf{SGX PHR}};

    \node[attacker, below=1cm of att.south] (modext) {};
    \node[below=1.0cm of att.south]           (innertitle) {\textbf{Shadow Tree}};
    \node[treenode, below=1.5cm of att.south] {}
      child {node[treenode] {}
        child {node[treenode] {}}
        child {node[leafnode] {...}}
      }
      child {node[treenode] {}};
    \node[below=.1cm of modext.south]        (title2) {\textbf{Attack Logic}};

    \node[right=2.0cm of att.east]              (title3) {\textbf{TEE}};
    \node[victim, below=.1cm of title3.south]   (tee) {};
    \node[below=.1cm of title3.south]           (title4) {\textbf{DT inference}};
    \node[treenode, below=.1cm of title4.south] {}
      child {node[treenode] {}
        child {node[treenode] {}}
        child {node[treenode] {}}
      }
      child {node[treenode] {}
        child {node[treenode] {}}
        child {node[treenode] {}}
      };

    \draw[arrow] (modext.west) -- +(-0.75, 0) |- node[left, pos=.25, text width=35pt] {1. Input or \texttt{finish}} (att.west);
    \draw[arrow] ([xshift=-10pt]att.south) -- node[left, pos=.5, text width=40pt] {5. Branch info., result} ([xshift=-10pt]modext.north);

    \draw[arrow] ([yshift=5pt]att.east) -| node[above, pos=.25] {2. Send input} ([xshift=-20pt]tee.north);
    \node[left=.35cm of title4] (tee1) {};
    \node[left=.95cm of tee1] (att1) {};
    \draw[arrow] ([yshift=15pt]tee) -| node[above, pos=0.25] {4. Label} ([xshift=20pt]att.south);

    \draw[arrow, dashed, color=orange] ([yshift=5pt]tee) -| node[below, pos=.5, xshift=25pt]{3. Execution leakage}([xshift=10pt]att.south);
    
\end{tikzpicture}